\documentclass{moriond}
\usepackage{lmodern}
\usepackage{calc}
\usepackage{graphicx}
\usepackage{booktabs}
\usepackage{textcomp}
\usepackage{xspace}
\usepackage{relsize}
\usepackage{amssymb}
\usepackage{amsmath}
\usepackage{listings}
\usepackage{microtype}
\usepackage{multirow}
\usepackage{tabularx}
\usepackage{array}
\usepackage{placeins}
\usepackage{cuted}
\usepackage{soul} 
\usepackage{fixltx2e}
\usepackage{slashed}

\def\be{\begin{equation}}
\def\ee{\end{equation}}
\def\bea{\begin{eqnarray}}
\def\eea{\end{eqnarray}}

\newcommand{\Photo}{\includegraphics[height=35mm]{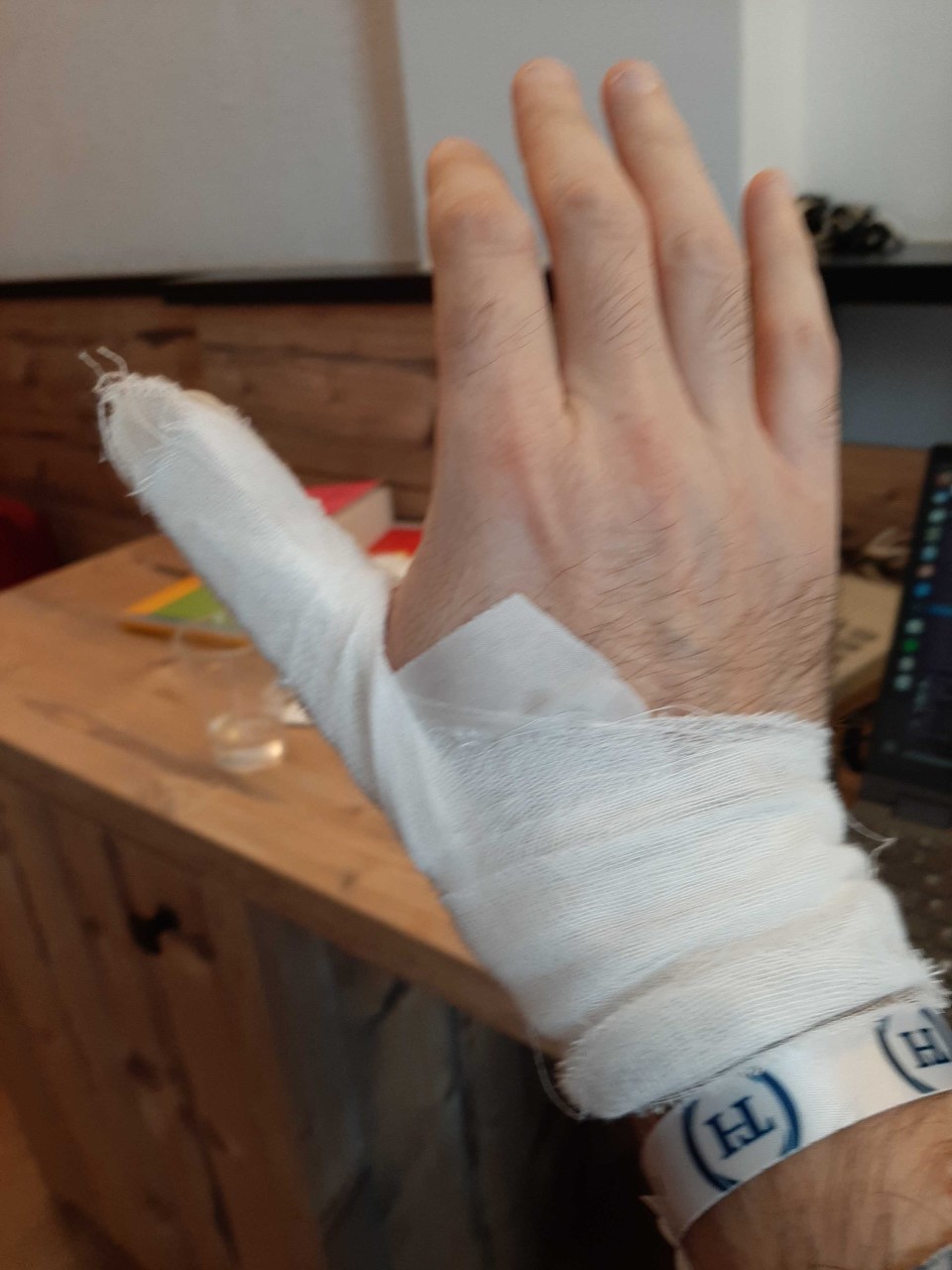}}
\newcommand{\La}{{\rm \Lambda}}

\newcommand{\Q}[2]{
  \if\relax\detokenize{#2}\relax
    \mathcal{Q}_{#1}
  \else
    \mathcal{Q}_{#1}^{(#2)}
  \fi
}

\newcommand{\C}[2]{
  \if\relax\detokenize{#2}\relax
    \mathcal{C}_{#1}
  \else
    \mathcal{C}_{#1}^{(#2)}
  \fi
}

\newcommand{\ddm}{\textsf{DirectDM}\xspace}

\begin{document}
\vspace*{4cm}

\title{Recent dark matter results from the GAMBIT collaboration}

\author{ M.J. WHITE on behalf of the GAMBIT Community}

\address{School of Physical Sciences, University of Adelaide,\\
Adelaide SA 5034, Australia}

\maketitle\abstracts{
In this proceeding, I present the results of a recent global fit of an effective field theory of dark matter, performed using the Global and Modular Beyond-the-Standard Model Inference Tool (GAMBIT). A Dirac fermion dark matter candidate is assumed to interact only with Standard Model quarks and gluons, through a general set of operators up to mass dimension 7. The results provide a reasonably up-to-date summary of our knowledge of possible WIMP interactions.}

\section{Introduction}
The nature of dark matter remains a mystery, despite abundant observational evidence for its gravitational interactions. The hypothesis that dark matter is comprised of weakly-interacting massive particles (WIMPs) continues to receive much attention, partly due to the fact that it can naturally explain the cosmologically-observed abundance, and partly because it is amenable to several experimental probes in the near future. There are a number of UV-complete explanations of WIMP dark matter, but it is useful to develop relatively agnostic ways of comparing constraints from cosmological, astrophysical and particle physics measurements.

A popular approach for doing so has long been to construct effective field theories (EFTs) of WIMP candidates interacting with Standard Model (SM) particles. Such theories are valid at energy scales well below a cut-off scale $\La$ at which the UV-physics is expected to be resolved, and one can proceed by enumerating all allowed higher dimensional operators that extend the SM with SM-DM interactions. This has the advantage of making minimal assumptions about the UV physics and, moreover, is accurate for low velocity environments such as direct and indirect DM detection experiments. Once one considers higher energy DM probes such as the ATLAS and CMS experiments of the Large Hadron Collider (LHC), however, a detailed investigation of the EFT validity is essential.

In this proceeding, I summarise the results of a recent global fit of a DM EFT, obtained using the Global-and-Modular beyond-Standard Model Inference Tool (GAMBIT)~\cite{GAMBIT:2021rlp,GAMBIT:2017yxo}. The open-source GAMBIT code is divided into a series of modules that implement statistics calculations and sampling algorithms~\cite{Martinez:2017lzg}, theoretical physics calculations~\cite{GAMBITModelsWorkgroup:2017ilg}, and experimental likelihoods from particle physics, astrophysics and cosmology experiments~\cite{GAMBIT:2017qxg,GAMBITDarkMatterWorkgroup:2017fax,GAMBITFlavourWorkgroup:2017dbx,GAMBITCosmologyWorkgroup:2020htv}. Recent innovations include a new system for automatically generating code from the Lagrangian density of a beyond-Standard Model physics theory~\cite{Bloor:2021gtp}. A recent survey of GAMBIT physics results can be found in Ref.~\cite{Kvellestad:2019vxm}.

\section{EFT details}
The DM EFT considered in the GAMBIT study is built on the assumption that the WIMP, $\chi$, is a Dirac fermion gauge-singlet. The interaction Lagrangian for the theory can be written as

\begin{equation}
  \mathcal{L}_{\rm{int}} = \sum_{a,d} \dfrac{\C{a}{d}}{\La^{d-4}} \Q{a}{d}\,,
\end{equation}
where $\Q{a}{d}$ is the DM-SM operator, $d\geq 5$ is the mass dimension of the operator, $\C{a}{d}$ is the
dimensionless Wilson coefficient associated to $\Q{a}{d}$ and $\La$ is the scale of new physics, which can be approximately identified with the mass of the mediator of the DM-SM interactions. The full Lagrangian density for the theory is
\begin{equation}
  \mathcal{L} = \mathcal{L}_{\rm{SM}} + \mathcal{L}_{\rm{int}} + \overline{\chi}\left(i\slashed{\partial}-m_\chi\right)\chi\,.
\end{equation}
Hence, the free parameters of the theory are the Wilson coefficients $\{ \C{a}{d} \}$, the WIMP mass $m_\chi$ and the scale of new physics $\La$. Note that, below the weak scale, the appropriate EFT has the Higgs, $W$, $Z$ and top quark integrated out, whilst that above the electroweak scale includes them, and the two EFTs must be appropriately matched in our study given that we apply both low energy and high energy constraints. For sufficiently large $\La$, the low-energy phenomenology is dominated by the lowest dimension operators, and we thus limit the study to the case $d\le 7$. Even then, the full space of operators is too large to realistically be explored in a global fit, and we make various simplifying assumptions. These include dropping dimension 6 operators that feature leptons (instead only including those with quarks), dropping operators that are products of DM and Higgs currents and dropping dimension 7 operators that have derivatives acting on DM fields. A full list of the operators included in the study (and the justifications for the omission of others) is provided in the original paper~\cite{GAMBIT:2021rlp}. Our code includes various effects arising from the renormalisation group running of the Wilson coefficients, including the mixing of different operators and threshold corrections that occur when the energy scale drops below the mass of one of the quarks. Our analysis makes use of the code \ddm \textsf{v2.2.0} \cite{Bishara:2017nnn,Brod:2017bsw}. 

\section{Scan details}

We perform various scans using the differential evolution scanner \textsf{Diver v1.4.0} \cite{Martinez:2017lzg}. In order to be able to neglect QCD resonances in the process $\chi \bar{\chi} \to q \bar{q}$, we restrict ourselves to $m_\chi > 5\,\mathrm{GeV}$. We also require $\La > 20\,\mathrm{GeV}$ in order to have a large separation of scales between $\La$ and the hadronic scale. We furthermore impose the bound $|\C{a}{d}| < 4\pi$ on all Wilson coefficients, and require $\La > 2 m_\chi$ in order to ensure that the EFT is valid for calculations of the WIMP relic abundance. Different upper bounds on $m_\chi$ and $\La$ are applied for different scans.

A long list of experimental constraints is applied, including:

\begin{itemize}
  \item direct dark matter detection constraints from the most recent XENON1T analysis, LUX 2016, PandaX 2016 and 2017 analyses, CDMSlite, CRESST-II and CRESST-III, PICO-60 2017 and 2019, and DarkSide-50. Our scans include nuisance parameters for the relevant astrophysical and nuclear uncertainties;
  \item the Planck measurement of the DM relic abundance;
\item the \texttt{Pass-8} combined analysis of 15 dSphs after 6 years of \emph{Fermi}-LAT data, targetting gamma ray production via dark matter annihilation;
\item constraints on the solar capture of dark matter based on the 79-string IceCube search for DM annihilation in the Sun;
\item cosmological constraints on DM annihilation in the early universe derived from the \emph{Planck} TT,TE,EE+lowE+lensing likelihoods, as well as the BAO data of 6dF, SDSS DR7 MGS, and the SDSS BOSS DR12 galaxy sample;
 \item searches for monojet events performed by the ATLAS and CMS experiments. The validity of the EFT is taken into account by modifying the missing energy distribution for $\slashed{E}_T > \La$ with either a hard cut-off or a continuous form factor suppression. Different scans are performed for different assumed conditions.
  
\end{itemize}

\section{Results}
\begin{figure*}[t]
        \centering
        \includegraphics[width=0.5\columnwidth]{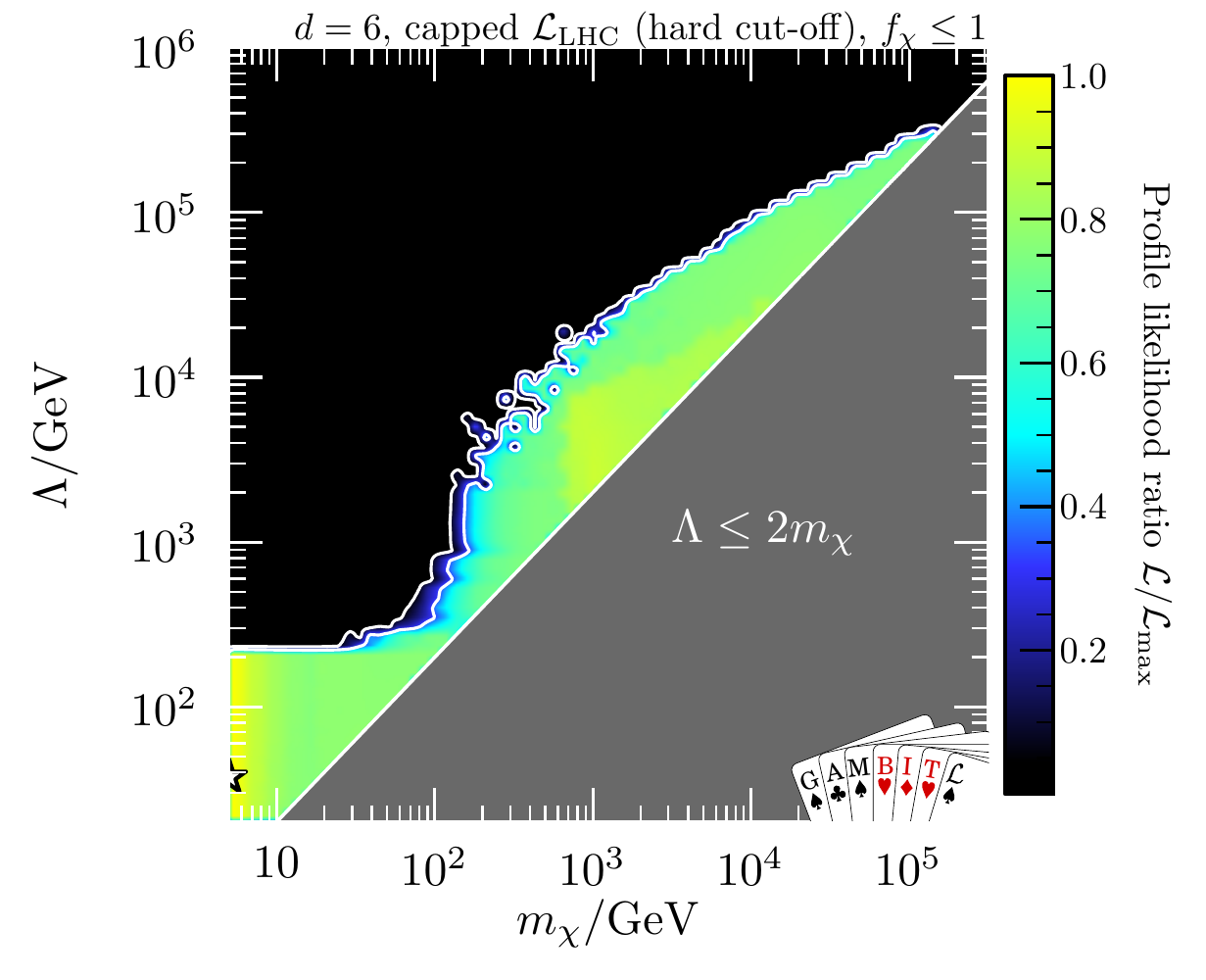}
        \caption{Profile likelihood in the $m_\chi$--$\La$ plane when considering only dimension-6 operators and capping the LHC likelihood at the value of the background-only hypothesis. The white contours indicate the $1\sigma$ and $2\sigma$ confidence regions and the best-fit point is indicated by the white star. For details see text.}
        \label{fig:dim_6_capped_main}
\end{figure*}

In Figure~\ref{fig:dim_6_capped_main}, we show the profile likelihood distribution in the $m_\chi$--$\La$ plane for a scan that only includes dimension-6 operators, and which assumes that the LHC likelihood is capped at the value of the background-only hypothesis (which tests the exclusion potential of the LHC monojet searches). The shaded region is excluded by the EFT validity condition, whilst the white star shows the best fit point. There is much viable paramter space, and the key features of the plot can be understood by thinking about three different qualitative regions of $\La$. For the highest $\La$ values, the EFT is valid for all experiments, but most experiments are insensitive to the DM physics. The constraints in this region are driven by the relic density requirement and its interplay with the perturbativity bound on the Wilson coefficients. For lower $\La$ values (comparable to LHC energies), there are strong LHC constraints on light WIMPs, resulting from the ATLAS and CMS monojet exclusions. For even lower $\La$ values, the EFT becomes invalid at the LHC, and the LHC constraints therefore disappear by construction. This has an interesting implication. For a WIMP mass of less than approximately 100 GeV, $\La \lesssim 200 \, \mathrm{GeV}$, meaning that the LHC should be sensitive to the UV physics, and one should hope to observe direct evidence of the DM-SM mediator.  

\section{Conclusions}

The GAMBIT community have performed a global fit of a dark matter effective field theory. Assuming that DM consists of a Dirac fermion interacting with SM quarks and gluons, much viable parameter space remains, and there are interesting experimental implications for the next decade of collider and dark matter direct and indirect search experiments.

\section*{Acknowledgments}
MJW wishes to thank the organising committee for the invitation to give a talk, and the conference secretaries for assistance when he broke his thumb on the ski slope! He is supported by the Australian Research Council grants CE200100008, DP180102209 and DP220100007. 

\section*{References}
\bibliography{mjw-moriond}
\end{document}